\documentclass[12pt]{article}

\usepackage[latin1]{inputenc}
\usepackage[T1]{fontenc}

\usepackage{amsmath}
\usepackage{physics}

\usepackage{enumitem}
\usepackage{float}
\usepackage{graphicx}

\usepackage{color}
\usepackage{dsfont}
\usepackage{ragged2e}
\usepackage{soul}
\usepackage[table]{xcolor}
\usepackage{makecell}

\usepackage{cite}
\usepackage{hyperref}
\usepackage{lipsum}

\def\beq{\begin{equation}}
\def\eeq{\end{equation}}
\def\bea{\begin{eqnarray}}
\def\eea{\end{eqnarray}}
\def\nn{\nonumber}
\def\roughly#1{\mathrel{\raise.3ex\hbox
{$#1$\kern-.75em\lower1ex\hbox{$\sim$}}}}

\def\order{\lower 1.8ex \hbox{\LARGE\~{}}}

\def\bsmumu{{\bar b} \to {\bar s} \mu^+ \mu^-}

\newcommand{\caret}{\textasciicircum}
\newcommand{\hc}{h.c.}

\begin{document}

\begin{flushright}
UdeM-GPP-TH-25-303 \\
LA-UR-25-21276
\end{flushright}

\begin{center}
\bigskip
\bigskip

{\Large \bf \boldmath The generic basis and flavour non-universal SMEFT} \\

\bigskip
\bigskip

\renewcommand*{\thefootnote}{\fnsymbol{footnote}}
{\large
Alakabha Datta $^{a,}$\footnote{datta@phy.olemiss.edu},
Jean-Fran{\c c}ois Fortin $^{b,}$\footnote{jean-francois.fortin@phy.ulaval.ca},
Jacky Kumar $^{c,}$\footnote{jacky.kumar@lanl.gov}, \\
David London $^{d,}$\footnote{london@lps.umontreal.ca},
Danny Marfatia $^{e,}$\footnote{dmarf8@hawaii.edu}
and Nicolas Sanfa{\c c}on $^{d,}$\footnote{nicolas.sanfacon@umontreal.ca}
}
\end{center}

\begin{flushleft}
~~~~~~~~~~~$a$: {\it Department of Physics and Astronomy,}\\
~~~~~~~~~~~~~~~{\it 108 Lewis Hall, University of Mississippi, Oxford, MS 38677, USA}\\
~~~~~~~~~~~$b$: {\it D\'epartement de Physique, de G\'enie Physique et d'Optique,}\\
~~~~~~~~~~~~~~~{\it Universit\'e Laval, Qu\'ebec, QC, Canada  G1V 0A6}\\
~~~~~~~~~~~$c$: {\it Theoretical Division, Los Alamos National Laboratory,}\\
~~~~~~~~~~~~~~~{\it Los Alamos, NM 87545, USA}\\
~~~~~~~~~~~$d$: {\it Physique des Particules, Universit\'e de Montr\'eal,}\\
~~~~~~~~~~~~~~~{\it 1375, ave Th\'er\`ese-Lavoie-Roux, Montr\'eal, QC, Canada H2V 0B3}\\
~~~~~~~~~~~$e$: {\it Department of Physics and Astronomy,}\\
~~~~~~~~~~~~~~~{\it University of Hawaii at Manoa, Honolulu, HI 96822, USA}\\
\end{flushleft}

\renewcommand*{\thefootnote}{\arabic{footnote}}
\setcounter{footnote}{0}

\begin{center}
\bigskip (\today)
\vskip0.5cm {\Large Abstract\\} \vskip3truemm
\end{center}

Whenever an anomaly in the flavour sector appears, analyses are performed examining whether it can be explained by adding a small number of carefully-chosen flavour non-universal four-fermion SMEFT operators. These analyses are typically carried out in the down or the up basis, i.e., it is assumed that the weak and mass eigenstates are the same for the left-handed down-type or up-type quarks. In these bases, there is no dependence on the matrices that transform from the weak to the mass basis, and which are unmeasurable in the Standard Model. In this paper, we argue that it is better to use a generic weak basis, in which no assumptions about the alignment of weak and mass eigenstates are made. The analysis now directly includes elements of the transformation matrices. By doing a fit to the data, it is possible to both determine if the flavour anomaly can be explained and extract the transformation matrices. In principle, this can be extended to a complete reconstruction of the Yukawa matrices.

\thispagestyle{empty}
\newpage
\setcounter{page}{1}
\baselineskip=14pt

\tableofcontents

\section{Introduction}

Although the standard model (SM) of particle physics has been enormously successful in describing the physics up to energy scales of $O({\rm TeV})$, it is not complete: it cannot explain a number of observations (e.g., neutrino masses, dark matter, the baryon asymmetry of the universe, etc.). We therefore conclude that there must be physics beyond the SM.

The LHC has searched for new particles for many years, but none have been observed. This implies that this new physics (NP), whatever it is, must be heavy (or extremely weakly coupled). One approach to analyzing such NP is to build models. Another approach uses model-independent effective field theories (EFTs). When the NP is integrated out, one obtains the Standard Model EFT, SMEFT \cite{Buchmuller:1985jz, Grzadkowski:2010es, Brivio:2017vri}. This EFT obeys the SM gauge symmetry, $SU(3)_C \times SU(2)_L \times U(1)_Y$ and contains only SM particles. The leading-order (dimension-4) terms in SMEFT are those of the SM; higher-order terms are suppressed by powers of the NP scale $\Lambda$ (i.e., the SM terms receive dimension-6 corrections).

The main purpose of this paper is to re-examine the way SMEFT is used to explain anomalies in the flavour sector. At present, there are a number of ``flavour anomalies,'' i.e., measurements of flavour-physics observables that disagree with the predictions of the SM, and there have been others in the past that have disappeared with new measurements. When such an anomaly appears, it is often noted that it can be explained by adding new SMEFT four-fermion operators. Since each fermion has three flavours and can be left-handed (LH) or right-handed (RH), there are many operators in a given class (four quarks, four leptons, two quarks and two leptons). Because of this, for practical reasons it is usually assumed that only a small number of carefully-chosen operators is present.

In addition, SMEFT operators are given in the weak basis, while the physical problem to be explained involves operators in the mass basis. It is therefore necessary to transform from the weak basis to the mass basis. For quarks, this involves transformation matrices in the LH and RH sectors, which introduce a number of additional unknown parameters. To avoid this problem, it is usually assumed that the weak and mass eigenstates are the same for the LH down-type quarks (the down basis) or the LH up-type quarks (the up basis). In this case, for the LH up-type (down-type) quarks, the transformation matrix from the down (up) weak basis to the mass basis is related to the (known) Cabibbo-Kobayashi-Maskawa (CKM) matrix, so the only unknown parameters are the coefficients of the SMEFT operators.

The point is that analyses that attempt to explain flavour anomalies generally make two assumptions:\footnote{Alternatively, assumptions such as minimal flavour violation (MFV) are made to restrict the flavour-violating operators \cite{DAmbrosio:2002vsn}.} (i) only a subset of SMEFT operators is present, and (ii) these operators are generated in the down or the up basis\footnote{Note, however,  that this choice is not stable under renormalization-group running. At the electroweak scale a \textit{back-rotation} has to be performed to return to the initial choice of basis \cite{Aebischer:2018bkb,Aebischer:2020lsx}.}  (for example, see Refs.~\cite{Aebischer:2015fzz, Bobeth:2017xry, Celis:2017doq, Cirigliano:2017tqn, Aebischer:2018csl, Crivellin:2022rhw, Greljo:2023bab, Dawid:2024wmp, Datta:2024zrl}). In this paper, we revisit these assumptions and suggest a better approach.

Indeed, instead of assuming that the operators are defined in the down or the up basis, we argue that it is better to use a generic weak basis, in which no assumptions about the alignment of the weak and mass eigenstates are made. It is true that this introduces additional unknown parameters into the analysis (the elements of the transformation matrices). However, the key point is that, when one does a fit to all the relevant data (including the flavour anomaly), there are enough observables that all parameters can be determined. That is, one can ascertain if this subset of SMEFT operators can explain the anomaly {\it and} one can extract some (or all) of the transformation matrices (and thus possibly the Yukawa textures). In other words, there is no need to make an assumption about the basis -- the data will tell us what the answer is.

In the SM, these individual transformation matrices are unmeasurable. It is only the CKM matrix, which is the product of two such matrices, that can be measured. This is because the couplings are universal, i.e., the fermions of all generations have the same $SU(2)_L \times U(1)_Y$ charges. On the other hand, in SMEFT, flavour non-universal couplings are permitted, and these can produce flavour anomalies. When one has only a subset of SMEFT operators of a given type, the extraction of elements of these matrices from the data is possible.

Note that there have been previous papers (e.g., see Ref.~\cite{Bhattacharya:2014wla}) that searched for NP explanations of flavour anomalies and carried out the analysis in the generic basis (though the language of SMEFT was not used). However, in the analysis of Ref.~\cite{Bhattacharya:2014wla}, certain assumptions were made about the elements of the transformation matrices in order to obtain a viable explanation; it was not realized that, in fact, these elements can be extracted from the data.

As is well known, if a good fit is found, one must then address the question: how is it that only this particular subset of operators is generated? This can only happen due to some symmetry in the underlying NP, which then restricts the type of NP models. We argue that this points to models in which there is a new gauge symmetry, and the NP particles are gauge bosons. That is, even though SMEFT analyses are meant to be model-independent, once one assumes that only a subset of operators is present, this necessarily has implications for the NP. (This also raises a second question: could these operators really be produced in the down or up bases when the NP is integrated out?) Note that this also applies to analyses that consider subsets of operators in order to put constraints on them.

To summarize the above discussion: if one attempts to explain a flavour anomaly by assuming a subset of SMEFT operators, it is also important to imagine a NP model (or models) that produces only these operators. In addition, one should not assume that these operators are defined in the down or up basis. Instead, one should take them to be in a generic weak basis -- by doing a fit to the data, one can identify this basis (which could turn out to be the down or up basis). We begin in Sec.~\ref{sec:known} with a review of the properties of SMEFT relevant to our arguments. Sec.~\ref{sec:new} examines which types of NP models can generate subsets of SMEFT operators, and contains a detailed discussion of the generic basis and its advantages. We conclude in Sec.~\ref{sec:conc}.

\section{Preliminaries: weak bases and subsets of operators}
\label{sec:known}

In this section, we review some well-known properties of SMEFT. We will use these to fix the notation and definitions that will be used in the next section. 

\subsection{\texorpdfstring{{\boldmath $[U(3)]^5$}}{[U(3)]\caret 5} transformations}

As SMEFT is valid above the weak scale, its operators are usually defined in the weak basis. That is, they are written as functions of the fermions $q_i^0$ (LH quark doublets), $\ell_i^0$ (LH lepton doublets), $u_i^0$, $d_i^0$ and $e_i^0$ (all three RH singlets), $i=1,2,3$. (Here, LH and RH denote left- and right-handed, and the superscript 0 indicates a generic weak basis.) One has the freedom to perform $[U(3)]^5$ transformations on these fields, where one $U(3)$ mixes the three generations of $q_i^0$ among themselves, and the other $U(3)$s act similarly on $\ell_i^0$, $u_i^0$, $d_i^0$ and $e_i^0$ \cite{Faroughy:2020ina}.

To see how this works, consider the semileptonic SMEFT operator\footnote{All the ${\cal O}_a$ up to dimension 6 were derived in Ref.~\cite{Grzadkowski:2010es}; this set of operators is known as the Warsaw basis.}
\beq
{\cal O}^{(1)}_{\ell q} = C^{(1)}_{\ell q}\vert_{prst} \, 
(\bar{\ell}_p^0 \gamma_{\mu} \ell_r^0) \, (\bar{q}_s^0 \gamma^{\mu} q_t^0) ~.
\label{SMEFTop}
\eeq
To simplify things, we focus only on transformations of the quark fields. For a given $(p,r)$, the Hamiltonian is
\beq
{\cal H} = \sum_{s,t} C^{(1)}_{\ell q}\vert_{prst} \, 
(\bar{\ell}_p^0 \gamma_{\mu} \ell_r^0) \, (\bar{q}_s^0 \gamma^{\mu} q_t^0) ~.
\label{H_orig}
\eeq
This corresponds to nine operators ($s,t=1,2,3$). We now perform a $U(3)$ transformation, i.e., a change of basis, on the LH quark fields: 
\beq
\tilde{q}^0 = S_L^\dagger \, q^0 ~.
\eeq
The Hamiltonian becomes 
\bea
{\cal H} &=& \sum_{s,t} C^{(1)}_{\ell q}\vert_{prst} \, 
(\bar{\ell}_p^0 \gamma_{\mu} \ell_r^0) \, (\bar{\tilde{q}}_i^0 \, (S_L^\dagger)_{is} \, \gamma^{\mu} \, (S_L)_{tj} \, \tilde{q}_j^0) \nn\\
&=& \sum_{i,j} {\tilde C}^{(1)}_{\ell q}\vert_{prij} \, 
(\bar{\ell}_p^0 \gamma_{\mu} \ell_r^0) \, (\bar{\tilde{q}}_i^0 \gamma^{\mu} \tilde{q}_j^0) ~,
\label{Htrans}
\eea
where
\beq
{\tilde C}^{(1)}_{\ell q}\vert_{prij} = (S_L^\dagger)_{is} \, C^{(1)}_{\ell q}\vert_{prst} \, (S_L)_{tj} ~.
\label{coeff_trans}
\eeq
This shows that, when a $U(3)$ transformation is applied, both the operators and Wilson coefficients are affected.

The above calculation can be restated very simply in a more familiar language. The Hamiltonian ${\cal H}$ can be viewed as a physical vector, for which the SMEFT operators are the basis vectors and the Wilson coefficients are the components. A change of basis modifies both the basis vectors and the components, but not the physical vector. Starting with a Hamiltonian ${\cal H}$ containing all operators, a typical SMEFT analysis would constrain ${\cal H}$ by fixing its \textit{a-priori} free components, the Wilson coefficients, to specific values in a given basis of the SMEFT operators. Clearly, if such an analysis could be done, it could also be performed in a different basis. This would lead to the same Hamiltonian, with the Wilson coefficients in the two bases related by Eq.~\eqref{coeff_trans}.

\subsection{WET and the down and up weak bases}

As noted in the Introduction, the dimension-4 operators in SMEFT are those of the SM. Higher-order, non-SM operators can provide new contributions to low-energy processes. Measurements of low-energy observables can therefore give information about the coefficients of these SMEFT operators. Below the weak scale, the physics is described by the Weak Effective Theory (WET), obtained by further integrating out the SM particles heavier than the $b$ quark. The WET operators obey the $SU(3)_C \times U(1)_{em}$ gauge symmetry. In order to get information about the SMEFT operators from WET observables, it is necessary to match SMEFT to WET.

One complication is that, while the SMEFT operators are defined in the weak basis, the WET operators are in the mass basis. It is therefore necessary to transform between the two bases. For quarks, this is done as follows. The (leading-order) Yukawa terms in SMEFT are as in the SM:
\beq
{\bar q}_i^0 \lambda^u_{ij} i \sigma_2 H^* u_j^0 + {\bar q}_i^0 \lambda^d_{ij} H d_j^0 + \hc
\eeq
When the Higgs gets a vacuum expectation value $v$, breaking the $SU(2)_L \times U(1)_Y$  symmetry to $U(1)_{em}$, this leads to the up- and down-quark mass matrices $M_u \equiv \lambda^u v$ and $M_d \equiv \lambda^d v$. Each of these is diagonalized by a bi-unitary transformation. For $M_u$, we have $(S_L^u)^\dag M_u S_R^u = M_u^\text{diag}$. This implies that $U_L^0 = S_L^u U_L$ and $U_R^0 = S_R^u U_R$, where $U \equiv (u,c,t)^T$ and the absence of superscripts indicates the mass basis. Similarly, $(S_L^d)^\dag M_d S_R^d = M_d^\text{diag}$, so that $D_L^0 = S_L^d D_L$ and $D_R^0 = S_R^d D_R$, where $D \equiv (d,s,b)^T$. The Cabibbo-Kobayashi-Maskawa (CKM) matrix is $V = (S_L^u)^\dag S_L^d$. The diagonalization of both mass matrices can only be done once the electroweak symmetry is broken.

Above the weak scale, where SMEFT is valid, we can use the $[U(3)]^5$ freedom to diagonalize either $\lambda^d$ or $\lambda^u$. We begin with a ``generic'' SMEFT Hamiltonian in which neither of the Yukawa coupling matrices is diagonal. We can use one $U(3)$ transformation to produce $U_R^0 = S_R^u U_R$ and another for $D_R^0 = S_R^d D_R$. The RH quarks in the SMEFT Hamiltonian are now in the mass basis. At the SMEFT level, we can now use another $U(3)$ to transform $q^0=S_L^d q^{\prime 0}$. This implies that the down-quark Yukawa term becomes ${\bar q}_i^0 \lambda^d_{ij} H d_j^0={\bar q}_i^{\prime 0}((S_L^d)^\dag)_{ik} \lambda_{kl}^d (S_R^d)_{lj} H d_j = {\bar q}_i^{\prime 0}\lambda_{ij}^{d,\text{diag}} H d_j$. That is, the down-quark Yukawa coupling matrix has been diagonalized. The upshot is that the LH down-quark weak eigenstates are now the same as the mass eigenstates. In summary, we have
\beq
q^{\prime 0} = \begin{pmatrix} u_L^{\prime 0} \\ d_L^{\prime 0} \end{pmatrix} = 
(S_L^d)^\dagger\begin{pmatrix} u_L^0 \\ d_L^0 \end{pmatrix}=\begin{pmatrix} V^\dag\, u_L \\ d_L \end{pmatrix} ~.
\label{eq:downbasis}
\eeq

Alternatively, we can transform $q^0=S_L^u q^{\prime\prime 0}$. This diagonalizes the up-quark Yukawa coupling matrix, so that the LH up-quark weak eigenstates are the same as the mass eigenstates. In this case, we have
\beq
q^{\prime\prime 0}=\begin{pmatrix} u_L^{\prime\prime 0} \\ d_L^{\prime\prime 0} \end{pmatrix}=(S_L^u)^\dagger\begin{pmatrix} u_L^0 \\ d_L^0 \end{pmatrix} = \begin{pmatrix} u_L \\ V\, d_L \end{pmatrix}  ~.
\label{eq:upbasis}
\eeq

Eqs.~(\ref{eq:downbasis}) and (\ref{eq:upbasis}) are referred to respectively as the ``down basis'' and ``up basis.''
Both of these bases are quite useful for matching SMEFT to WET, because there is no dependence on the (unknown) transformation matrices $S_L^{u,d}$ -- only the (known) CKM matrix is involved \cite{Aebischer:2015fzz, Aebischer:2017ugx, Jenkins:2017jig}. These bases are also convenient for performing renormalization-group running of the SMEFT Hamiltonian from the scale $\Lambda$ down to the electroweak scale. The reason is that the Yukawa couplings, which figure in the calculation of the running, are {\it a priori} unknown. However, in these two bases, one of the Yukawa matrices is diagonal, i.e., it is proportional to the matrix of quark masses, while the other is a product of the other matrix of quark masses and the CKM matrix. In these bases, there are therefore no longer any unknown parameters coming from the $S_L^{u,d}$ matrices.

\subsection{Subsets of SMEFT operators}

Above, we examined what happens to the Wilson coefficients and the SMEFT operators when one performs a $[U(3)]^5$ transformation, i.e., a change of basis. We saw that, if the Hamiltonian $\cal{H}$ contains all of the contributing operators in one basis, a transformation to another basis will generally produce another Hamiltonian containing all of the contributing operators. The upshot is that a SMEFT analysis of the corresponding observables is basis independent. Although the values of the Wilson coefficients in different bases are different, they are related by the change of basis.

On the other hand, if only a subset of operators is used, the underlying basis must be specified, since the values of the Wilson coefficients are not basis-independent \cite{Silvestrini:2018dos, Aebischer:2020dsw}. For example, the physics described by the Hamiltonian of Eq.~\eqref{H_orig} with restrictions $C^{(1)}_{\ell q}\vert_{prst}=0$ unless $prst=2222$ or $2233$ is different from the physics described by the Hamiltonian of Eq.~\eqref{Htrans} with restrictions ${\tilde C}^{(1)}_{\ell q}\vert_{prij}=0$ unless $prij=2222$ or $2233$.

In fact, in the literature it is often assumed that only a subset of such operators is present. For this case, it is worthwhile to examine the basis dependence in more detail. Consider again the SMEFT operator ${\cal O}^{(1)}_{\ell q}$ of Eq.~(\ref{SMEFTop}), and suppose, as above, that the generation indices take only the two values $prst = 2222$ and $2233$, i.e., not all nine quark operators are present (and the leptonic indices are fixed). We are interested in the matching of these ${\cal O}^{(1)}_{\ell q}$ operators to WET. We have already described how the quarks transform between the weak and mass bases. For the leptons, we can again use one $U(3)$ transformation to produce $e_R^0 = S_R^e e_R$, so that the RH charged leptons in the SMEFT Hamiltonian are now in the mass basis. For the LH leptons, if we neglect neutrino masses, we can also take the LH charged-lepton fields in the weak basis to be the same as in the mass basis.

On the one hand, starting with these two SMEFT operators in the down basis (Eq.~\ref{eq:downbasis}), the WET operators involve only two down-quark transitions, but nine up-quark transitions:

\begin{equation}
\begin{aligned}
    \label{eq:two_ops_down_basis}
    & \sum_{k=2,3} {\tilde C}^{d,(1)}_{\ell q}\vert_{22kk} \, 
    (\bar{\ell}_2^0 \gamma_{\mu} \ell_2^0) \, (\bar{q}_k^{\prime 0} \gamma^{\mu} q_k^{\prime 0}) 
    \\
    &\qquad= \sum_{k=2,3} {\tilde C}^{d,(1)}_{\ell q}\vert_{22kk}\left[ \bar{\mu}_L \gamma_{\mu} \mu_L + \bar{\nu}_{\mu L} \gamma_{\mu} \nu_{\mu L} \right]
    \left[ \bar{d}_{k,L} \gamma^{\mu} d_{k,L} + \bar{u}_{i,L} V_{ik} \gamma^{\mu} V^\dag_{kj} u_{j,L} \right] ~.
\end{aligned}
\end{equation}
On the other hand, starting with the same two SMEFT operators in the up basis (Eq.~\ref{eq:upbasis}), the WET operators involve nine down-quark transitions, but only two up-quark transitions:
\begin{equation}
\begin{aligned}
    \label{eq:two_ops_up_basis}
    & \sum_{k=2,3} {\tilde C}^{u,(1)}_{\ell q}\vert_{22kk} \, 
    (\bar{\ell}_2^0 \gamma_{\mu} \ell_2^0) \, (\bar{q}_k^{\prime\prime 0} \gamma^{\mu} q_k^{\prime\prime 0}) 
    \\
    &\qquad= \sum_{k=2,3} {\tilde C}^{u,(1)}_{\ell q}\vert_{22kk}\left[ \bar{\mu}_L \gamma_{\mu} \mu_L + \bar{\nu}_{\mu L} \gamma_{\mu} \nu_{\mu L} \right]
    \left[ \bar{d}_{i,L} V^\dag_{ik} \gamma^{\mu} V_{kj} d_{j,L} + \bar{u}_{k,L} \gamma^{\mu} u_{k,L} \right] ~.
\end{aligned}
\end{equation}
This shows that the predictions in the two bases are very different.

\subsection{Subsets of operators from underlying new physics}
\label{sec:subsets_of_operators_from_NP}

In the previous subsection, we discussed characteristics of subsets of operators, but no explanation was given for the existence of this particular subset. This explanation must come from the underlying NP, and usually arises from symmetry arguments. 

For example, consider again the case discussed in the previous section: we take the Hamiltonian of Eq.~\eqref{H_orig} with restrictions $C^{(1)}_{\ell q}\vert_{prst}=0$ unless $prst=2222$ or $2233$. Can we find a NP model that explains why only two operators are nonzero, while the seven others are absent? One such model could be an extension of the SM gauge symmetry by a $U(1)'$ symmetry at the TeV scale. The gauge boson associated with this new symmetry is a $Z'$. It can be chosen to couple only to LH SM fermions. The couplings are
\begin{equation} \label{eq:Zprime_lagrangian}
-\mathcal{L}_{Z'}  \supset
g' \left[ C_\ell^i \bar \ell_i^0 \gamma^\mu \ell_i^0 + C_q^j  \bar q_j^0 \gamma^\mu q_j^0 \right] Z^\prime_\mu ~,                                   
\end{equation}
where $C_\ell^i$ and $C_q^j$ are respectively the $U(1)'$ charges for the $i^{th}$ and $j^{th}$ generations of leptons and quarks. When the $Z'$ is integrated out at scale $\Lambda = M_{Z'}$, three types of four-fermion operators are generated at tree level. There are those with four leptons, four quarks, and two leptons and two quarks: 
\beq
\label{eq:Zp_SMEFT_operators}
-\frac{g^{\prime2}}{2 M_{Z'}^2} \left[
C_\ell^i C_\ell^j (\bar{\ell}_i^0 \gamma^{\mu} \ell_i^0) (\bar{\ell}_j^0 \gamma_{\mu} \ell_j^0)
+ C_q^i C_q^j (\bar{q}_i^0 \gamma^{\mu} q_i^0) (\bar{q}_j^0 \gamma_{\mu} q_j^0)
+ 2 C_\ell^i C_q^j (\bar{\ell}_i^0 \gamma^{\mu} \ell_i^0) (\bar{q}_j^0 \gamma_{\mu} q_j^0) \right] ~.
\eeq
If only $ C_\ell^2$, $C_q^2$ and $C_q^3$ are nonzero, and $C_q^2 \ne C_q^3$, the only ${\cal O}^{(1)}_{\ell q}$ operators produced have indices 2222 and 2233.\footnote{We note that this model is not complete.  Apart from the usual consistency constraints (e.g., anomaly cancellations), some of the Yukawa couplings are forbidden. This can be corrected by introducing several Higgs fields. Since these details are not necessary to illustrate our point, we do not pursue the construction of a full model here.}

\section{Analyzing subsets of operators}
\label{sec:new}

It is a common practice to examine whether a particular flavour anomaly can be explained by adding certain SMEFT four-fermion operators. Indeed, there are many papers in the literature in which a subset of operators is assumed to be present (for example, see Refs.~\cite{Aebischer:2015fzz, Bobeth:2017xry, Celis:2017doq, Cirigliano:2017tqn, Aebischer:2018csl, Crivellin:2022rhw, Greljo:2023bab, Dawid:2024wmp, Datta:2024zrl}). These analyses are carried out in the down or up basis. This is because the unknown parameters in the $S_L^{u,d}$ transformation matrices do not appear in these bases. In this section, we propose an alternative approach.   As we will show, it is not necessary to make any assumptions about the basis -- under certain conditions, this information can be obtained from the data.

\subsection{Underlying NP models}

It is known that a subset of operators can only be produced through a symmetry in the underlying NP. What kinds of NP models possess such a symmetry?

If the NP particles have spin 1, they are generally gauge bosons associated with a new gauge symmetry. By assigning the quarks and leptons flavour non-universal charges under this new symmetry, one can produce subsets of operators. One example of such a model was given in Sec.~\ref{sec:subsets_of_operators_from_NP}.

Furthermore, in general there are a variety of different NP models that can generate this subset of SMEFT operators. And these different models may lead to different effects below the NP scale. For example, in the the $U(1)'$ model of Sec.~\ref{sec:subsets_of_operators_from_NP}, not only is the desired subset of semileptonic SMEFT operators generated, other fully-leptonic and fully-hadronic operators are also produced. Which additional operators are produced can vary from model to model.

On the other hand, if the NP particles have spin 0, such as in the 2-Higgs-doublet model, then the problem is that, in general, such scalar models do not contain a symmetry that can be used to explain why only a subset of SMEFT operators is nonzero. In the context of such NP models, this choice of operators is completely arbitrary, i.e., it is not very well motivated.

The point is that, if only a subset of operators is assumed, this is most likely the result of a gauge symmetry with flavour non-universal charges. This subset is produced when the NP is integrated out. Note that we have not yet determined in what basis this subset of operators is produced.

\subsection{\boldmath The generic basis}
\label{sec:generic_basis}

Suppose the measurement of an observable disagrees with the prediction of the SM. The obvious question is then: what type of NP can explain this anomaly? Here it is common to assume a subset of SMEFT operators (usually one or two) to see if this can account for the observed deviation from the SM (for example, see Refs.~\cite{Aebischer:2015fzz, Bobeth:2017xry, Celis:2017doq, Cirigliano:2017tqn, Aebischer:2018csl, Crivellin:2022rhw, Greljo:2023bab, Dawid:2024wmp, Datta:2024zrl}). 

As an example, consider a flavour anomaly from the past. Before 2022, there were a number of measurements of observables that suggested that there was NP in the transition $\bsmumu$. These included a variety of observables in decays dominated by $\bsmumu$. In addition, the measurements of 
$R_{K^{(*)}}$, where

\beq
R_{K^{(*)}} = \frac{{\cal B} (B \to K^{(*)} \mu^+ \mu^-)}{{\cal B} (B \to K^{(*)} e^+ e^-)} ~,
\eeq
disagreed with the SM: the SM predicts $R_{K^{(*)}} \simeq 1$, but the measured values differed from unity by $\simeq 2.5\sigma$. (For a review of the $\bsmumu$ anomaly as of 2021, see Ref.~\cite{London:2021lfn}.) This anomaly was greatly studied at the time, and one of the most promising explanations was that there is a NP contribution to the WET operator $({\bar\mu}_L \gamma^\mu \mu_L)({\bar b}_L \gamma_\mu s_L)$.

Now, suppose we take the scenario described in Sec.~\ref{sec:subsets_of_operators_from_NP}: only the semileptonic ${\cal O}^{(1)}_{\ell q}$ operators with indices 2222 and 2233 are present. That is, we add only a subset of all possible ${\cal O}^{(1)}_{\ell q}$ operators. How would we determine if the addition of this subset of SMEFT operators could explain the $\bsmumu$ anomaly?

Such analyses are usually done in the down or up basis. As explained above, this is usually done for convenience -- the transformation from the weak basis to the mass basis is simple. Suppose we work in the down basis. The methodology is as follows. When we run down to the WET scale using the renormalization group, the dominant operators will be the same operators as those at the NP scale. (Note that, if one does this in the context of the $Z'$ model, there are other dominant operators as can be seen in Eq.~\eqref{eq:Zp_SMEFT_operators}.) There are also subdominant operators of all types generated by the running. These include four-lepton, four-quark, and other semileptonic operators, etc. For all of these operators, we must transform from the down basis to the mass basis using Eq.~(\ref{eq:downbasis}). The end result is that there are many operators at the WET scale; all of their coefficients depend on the same quantities: $C^{(1)}_{\ell q}$ and the CKM matrix $V$. Since $V$ is known, there are two unknown parameters.  

A fit is now performed, ideally including all the observables to which these operators contribute. These include all the $\bsmumu$ observables and $R_{K^{(*)}}$. If a good fit is found, this tells us that the $\bsmumu$ anomaly can be explained by the addition of this subset of SMEFT operators in the down basis, with coefficients equal to the best-fit values of the $C^{(1)}_{\ell q}$. If a poor fit is found, then this scenario doesn't work.

There are two problems with this approach. Most importantly, it's not systematic. The down basis was assumed somewhat arbitrarily. If a good fit is found, that's great. But what's the next step if a poor fit is found? Do we try this subset of operators in another basis (e.g., the up basis)? Do we choose a different subset of operators? Furthermore, there is still the question of the underlying NP: could it really generate operators in the down basis when the heavy particles are integrated out?

The main point of this paper is that it is better to make no assumptions about the basis. This means that we should work in what we call the ``generic'' basis. This is defined as a weak basis in which there is no assumption about the alignment of Yukawa couplings. In the context of the underlying NP model, the generic basis is the weak basis in which the flavour non-universal charges are well-defined. Note that this does not preclude the operators being defined in the down (or up) basis. But if this is the case, we will get this information from the data.

In contradistinction with the down and up bases, the generic basis is simply
\bea
\label{eq:generic_basis}
q^0 =
\begin{pmatrix} u_L^0 \\ d_L^0 \end{pmatrix}=\begin{pmatrix} S_L^u \, u_L \\ S_L^d \, d_L \end{pmatrix}=S_L^d \,\begin{pmatrix} V^\dagger \, u_L \\ \, d_L \end{pmatrix} ~,
\eea
and the matching to WET is done via this equation. Note that, since the CKM matrix $V = (S_L^u)^\dag S_L^d$, we have expressed the change of basis in terms of $V$ and $S_L^d$. Thus, when we transform the two semileptonic operators to the WET basis, there are now nine down-quark transitions and nine up-quark transitions:
\begin{align}
    \label{eq:two_ops_generic_basis}
    \sum_{k=2,3} C^{(1)}_{\ell q}\vert_{22kk} \, 
    (\bar{\ell}_2^0 \gamma_{\mu} \ell_2^0) \, (\bar{q}_k^0 \gamma^{\mu} q_k^0) 
    \\
    \nonumber
    \qquad= \sum_{k=2,3} (S_L^d)^\dag_{ik} C^{(1)}_{\ell q}\vert_{22kk} (S_L^d)_{kj} &\left[ \bar{\mu}_L \gamma_{\mu} \mu_L + \bar{\nu}_{\mu L} \gamma_{\mu} \nu_{\mu L} \right] \left[\bar{d}_{i,L} \gamma^{\mu} d_{j,L} + \bar{u}_{a,L} V_{ai} \gamma^{\mu} (V)_{jb}^\dagger u_{b,L} \right] ~.
\end{align}
This is like a combination of Eqs. \eqref{eq:two_ops_down_basis} and \eqref{eq:two_ops_up_basis}. Here, the coefficients at the WET scale involve elements of the CKM matrix $V$ and the transformation matrix $S_L^d$.

To proceed with the analysis, we follow a similar methodology to that used in the down basis. We run down to the WET scale using the renormalization group, generating many operators. For all of these operators, we transform from the generic basis to the mass basis using Eq.~(\ref{eq:generic_basis}). We have the same operators as was found starting in the down basis, but here their coefficients depend on $C^{(1)}_{\ell q}$ (two real parameters), $V$ (three angles and one phase, after removing five unphysical phases) and $S_L^d$ (three angles and five phases, after removing the last unphysical phase). And since $V$ is known, here there are ten unknown parameters.

In order to extract these parameters, we must perform a fit with more than ten observables. Fortunately, when run down to the WET scale, these operators contribute to far more than ten observables. In addition to $R_{K^{(*)}}$, there are many observables measured in processes dominated by the $\bsmumu$ transition. These include the angular observables of $B^0 \to K^{*0} \mu^+ \mu^-$ (particularly $P_5^\prime$) and $B^0_s \to \phi \mu^+ \mu^-$, the differential branching ratios of $B^0 \to K^0 \mu^+ \mu^-$, $B^0_s \to \phi \mu^+ \mu^-$ and $B \to X_s \mu^+ \mu^-$, and $\mathcal{B}(B^0_s \to \mu^+ \mu^-)$. Other processes that would likely be impacted are $K^0$-$\bar{K}^0$, $B_d^0$-$\bar{B}^0_d$, $B_s^0$-$\bar{B}_s^0$ and $D^0$-$\bar{D}^0$ mixing \cite{Buras:2013ooa}, which provide many observables, both CP-conserving ($\Delta M_K$, $\Delta M_d$, $\Delta M_s$, $\Delta M_D$) and CP-violating ($K$ decays: $\varepsilon_K$, $\varepsilon^\prime / \varepsilon$; $B_d^0$ decays: $\alpha$, $\beta$, $\gamma$; $B_s^0$ decays: $\phi_s$; $D$ decays: $|q/p|_D$).
Finally, electroweak precision observables should also be included, in order to verify that the NP does not affect them significantly. For example, 17 such observables have been measured in decays of the $Z$ boson. Some of these ($\Gamma_Z$, $\sigma^0_\text{had}$, $R^0_\ell$ and $A^{0,\ell}_\text{FB}$) come from LEP \cite{Janot:2019oyi}, while others ($A_\ell, R^0_q$, $A^{0,q}_\text{FB}$ and $A_q$) come from both  LEP and SLC \cite{ALEPH:2005ab} ($\ell = e, \mu, \tau$ and $q = b, c$). In a similar vein, we can include the four $W$ observables $m_W$ \cite{Electroweak:2003ram} and $\Gamma(W \to \ell \nu)$ \cite{DELPHI:2003ftu}. 

If a good fit is found, we will have shown that this subset of operators can explain the $\bsmumu$ anomaly, with coefficients equal to the best-fit values of the $C^{(1)}_{\ell q}$. 
In addition, we will have reconstructed $S_L^d$ (and, by extension, $S_L^u$), which determines the basis in which these operators are defined. If it turns out that $S_L^d = \mathds{1}$ and $S_L^u = V^\dagger$, then we will know that we are in the down basis. And if $S_L^d = V$ and $S_L^u = \mathds{1}$, we will know that we are in the up basis. There is no need to assume this from the very beginning, as the data will tell us which basis it is. Moreover, it is likely that Nature will show that we are in neither the down or the up basis, and this information will have been lost by an initial assumption about the basis. 

As we have noted above, there must be a NP explanation for the presence of a particular subset of operators. And there are, in general, a number of different NP models that one can construct that generate this subset. The above analysis can be carried out in the context of any of these models (and one can even be agnostic and make no assumption about the model). What one finds for $S_L^d$ will depend on which underlying NP model is assumed. (This simply reflects the fact that, while all NP models generate operators in a weak basis when the NP is integrated out, the characteristics of this weak basis will vary from model to model.)

Finally, if a poor fit is found, we will know definitively that the anomaly cannot be explained by adding this subset of operators, irrespective of the basis.

\subsection{Analysis in the down basis}

In the above scenario, we have two semileptonic SMEFT operators in the generic basis. If we were to transform to the down basis, in general we would generate all possible semileptonic SMEFT operators. This might lead one to think that, if we did the analysis in the down basis, keeping all the semileptonic operators, we would obtain the same results as if we did the analysis in the generic basis.

However, as we show below, the analysis in the down basis is fundamentally different from the analysis in the generic basis. Most importantly, due to the parametrization, it is not possible to extract the transformation matrix $S_L^d$ when all the semileptonic SMEFT operators are turned on, simply because the transformation matrices are not part of the analysis. 

The fact that the analyses are not the same can be seen most easily by comparing the number of free parameters in the two scenarios. If all semileptonic SMEFT operators are kept, the analysis involves nine free real parameters in the Wilson coefficients $C_{\ell q}^{(1)}|_{22st}$, since the matrix of Wilson coefficients must be hermitian for the Lagrangian to be real. On the other hand, working in the generic basis, we have two real parameters for the two Wilson coefficients $C_{\ell q}^{(1)}|_{2222}$ and $C_{\ell q}^{(1)}|_{2233}$, and eight real parameters (three angles and five phases) from the transformation matrix $S_L^d$, for a total of ten free real parameters. Clearly, working with a subset of SMEFT operators in the generic basis is not equivalent to working in the down (or up) basis with all relevant SMEFT operators.

This point is made even stronger in the context of a NP model. In our $Z'$ example, in addition to the semileptonic SMEFT operators, four-lepton and four-quark SMEFT operators are also generated [Eq.~(\ref{eq:Zp_SMEFT_operators})].  From the generic basis point of view, the parameter counting stays the same: assuming the couplings $C_\ell^2$, $C_q^2$ and $C_q^3$ are known, there are the two real parameters $g'$ and $M_{Z'}$, and the three angles and five phases in $S_L^d$. But in the analysis in the down basis, in which all four-lepton, four-quark and semileptonic SMEFT operators are allowed, the number of free parameters increases drastically. This clearly demonstrates that our approach of working in the generic basis is not equivalent to simply working in the down (or up) basis with all relevant SMEFT operators.

\subsection{Textures of Yukawa couplings}

We have shown in Sec.~\ref{sec:generic_basis} that, in the context of our $Z'$ model with LH operators, it is possible, in principle, to extract the angles and phases of the $S_L^d$ transformation matrix from the data. This information, along with the transformation matrix $S_L^u$ obtained from the CKM matrix, tells us how to transform the LH quarks from the generic basis, where the flavour non-universal gauge charges are well-defined, to the mass basis. 

However, there is nothing inherently special about our choice of operators. Let us put aside the $\bar{b} \to \bar{s} \mu^+ \mu^-$ anomaly and suppose that, for some well-motivated reason, our model also contained operators with flavour non-universal couplings for RH quarks. The WET Wilson coefficients associated with these new operators would then be functions of the $S_R^u$ and $S_R^d$ transformation matrices. If we performed a fit with enough observables, we could measure -- again, in principle -- the angles and phases of $S_L^d$, $S_R^u$ and $S_R^d$, which means that we could fully reconstruct the quark Yukawa couplings in the generic basis.

As in the previous section, this can be shown by properly accounting for all the free parameters to be fitted. Prior to electroweak symmetry breaking, our free parameters lie in the Yukawa couplings $\lambda^u$ and $\lambda^d$, and in the SMEFT Wilson coefficients. Since the Yukawa couplings are arbitrary complex matrices, they each contain 18 real parameters, for a total of 36 real parameters. To that amount, we add the number of independent degrees of freedom in the Wilson coefficients.

After symmetry breaking, we transform to the mass basis using the $S_L^{u,d}$ and $S_R^{u,d}$ matrices. Each of them is unitary and thus contains nine real parameters (three angles and six phases). To that, we add the six components of the diagonalized Yukawa couplings, $\lambda^{u,\text{diag}}$ and $\lambda^{d,\text{diag}}$, for a total of 42 real parameters. Moreover, the SMEFT Wilson coefficients only change through the $S_L^{u,d}$ and $S_R^{u,d}$ transformation matrices such that the number of degrees of freedom related to them stays the same. 

Given the mismatch between the number of parameters before and after symmetry breaking, we know that there are six unphysical phases that can be removed by field redefinitions. As is done in the SM, five of these field redefinitions can be used to remove five phases from the CKM matrix, leaving only a single phase. The sixth redefinition removes a phase from $S_L^d$. To summarize, before electroweak symmetry breaking, the Yukawa matrices contain 36 real parameters. After symmetry breaking, these are transformed into six quark masses, three angles and one phase for the CKM matrix, three angles and five phases for $S_L^d$, and three angles and six phases for each of $S_R^u$ and $S_R^d$. This makes a total of 36 real parameters, as required, see Table \ref{tab:parameter_count}.

\begin{table}[H]
    \centering
    \begin{tabular}{|c|ccccccccc|}
        \hline & & & & & & & & & \\[-2mm]
         & $\lambda^u$ & $\lambda^d$ & $\Longleftrightarrow$ & $\lambda^{u,\text{diag}}$ & $\lambda^{d,\text{diag}}$ & $V$ & $S_L^d$ & $S_R^u$ & $S_R^d$ \\[3mm]
         \# Parameters & 18 & 18 & = & 3 & 3 & 4 & 8 & 9 & 9 \\
         \hline
    \end{tabular}
    \caption{The number of parameters before and after electroweak symmetry breaking.}
    \label{tab:parameter_count}
\end{table}

To perform a fit and reconstruct the textures of the Yukawa couplings, we would need to fix a total of 26 real unknown parameters (the 9 angles and 17 physical phases in $S_L^d$, $S_R^u$ and $S_R^d$), in addition to the degrees of freedom in the SMEFT Wilson coefficients. We would then require a set of at least as many observables that would be functions of the parameters we want to measure. Through such a procedure, measurements would dictate what are the best-fit Yukawa textures in the generic basis relevant to our NP model.  Obviously, similar considerations can be applied to the lepton sector, with or without neutrino masses.

\section{Conclusion}
\label{sec:conc}

Because we know that there must be physics beyond the SM, and since no new particles have been discovered at the LHC, much attention is paid to all indirect hints of NP. In particular, whenever an anomaly in the flavour sector appears, there is often a great deal of activity trying to ascertain which types of NP effects can account for the result. Explanations often involve the addition of four-fermion operators, and SMEFT is a useful tool to use, as it contains all possible operators that respect the $SU(3)_C \times SU(2)_L \times U(1)_Y$ symmetry of the SM. 

Suppose that one wants to examine whether a particular flavour anomaly can be explained by adding NP operators. Because there are many operators of a given class (four quarks, four leptons, semileptonic), for practical reasons it is usually assumed that only a subset of carefully-chosen flavour non-universal operators is present. A complication is that, in order to explain the anomaly, operators in the mass basis must be considered, but the SMEFT operators are defined in the weak basis. It is therefore necessary to use transformation matrices to pass from the weak to the mass basis, and this introduces new unknown parameters into the problem. (In the SM, these transformation matrices are unmeasurable; it is only the product of two of them, in the form of the CKM matrix $V$, that can be measured.) 

In order to deal with this, it is often assumed that the SMEFT operators are defined in the down or the up basis. In the down (up) basis, the weak eigenstates of the LH down-type (up-type) quarks are the same as the mass eigenstates. In these bases, the transformation matrices do not appear; only $V$ is present. The only unknown parameters are the coefficients of the SMEFT operators.

In this paper, we argue that important information is lost by working in the down or up basis, and we propose an alternative approach. It is not necessary to make assumptions about the alignment of the weak and mass eigenstates -- one can simply work in a generic weak basis. It is true that the transformation matrices appear in this basis, leading to additional unknown parameters. However, when a fit to the data is performed, there are enough observables that all parameters can be determined. That is, if a good fit is found, not only will this confirm that the chosen subset of operators can explain the anomaly, the transformation matrices will also be measured. There is no need to make an assumption about the basis in which the operators are defined. The data will provide this information (and it may turn out that it is the down or up basis).

The passage from the weak basis to the mass basis involves four transformation matrices $S_L^{d,u}$ and $S_R^{d,u}$, that act on the LH and RH down-type and up-type quarks. In the paper, we gave the example of a flavour anomaly that could be explained by adding operators that involved only LH quarks, and showed how $S_L^{d,u}$ could be extracted from the analysis. If another flavour anomaly could be explained by adding operators involving RH quarks, a similar analysis could be used to obtain $S_R^{d,u}$. In this case, we could completely reconstruct the down-quark and up-quark Yukawa matrices.

Finally, there is an additional question that should be addressed: how is it possible to generate only a particular subset of operators? This can only happen if the underlying NP model possesses a certain symmetry. In this paper, we argue that this points to models with an additional gauge symmetry, with flavour non-universal charges. In such models, it is unlikely that the operators are produced in the down or up basis when the NP particles are integrated out. It is a useful exercise to construct a NP model (or models) that generate this subset of operators, and repeat the above analysis, including the other operators produced by the model.

\addcontentsline{toc}{section}{References}
\small
\bibliographystyle{JHEP}
\bibliography{refs}

\end{document}